\documentclass[12pt]{iopart}
\usepackage{iopams}
\usepackage{graphicx}
\begin{document}

\title[PageRank and rank-reversal dependence on the damping factor]
{PageRank and rank-reversal dependence on the damping factor}

\author{Seung-Woo Son, Claire Christensen, Peter Grassberger and Maya Paczuski}
\address{Complexity Science Group, Department of Physics and
Astronomy, University of Calgary, Alberta, Canada}
\ead{swson@ucalgary.ca}

\begin{abstract}
PageRank (PR) is an algorithm originally developed by Google to
evaluate the importance of web pages. Considering how deeply
rooted Google's PR algorithm is to gathering relevant information
or to the success of modern businesses, the question of
rank-stability and choice of the damping factor (a parameter in
the algorithm) is clearly important. We investigate PR as a
function of the damping factor $d$ on a network obtained from a
domain of the World Wide Web, finding that rank-reversal happens
frequently over a broad range of PR (and of $d$). We use three
different correlation measures, Pearson, Spearman, and Kendall, to
study rank-reversal as $d$ changes, and show that the correlation
of PR vectors drops rapidly as $d$ changes from its frequently
cited value, $d_0=0.85$. Rank-reversal is also observed by
measuring the Spearman and Kendall rank correlation, which
evaluate relative ranks rather than absolute PR. Rank-reversal
happens not only in directed networks containing rank-sinks but
also in a single strongly connected component, which by definition
does not contain any sinks. We relate rank-reversals to
rank-pockets and bottlenecks in the directed network structure.
For the network studied, the relative rank is more stable by our
measures around $d=0.65$ than at $d=d_0$.
\end{abstract}

\maketitle

\section{Introduction}

Web pages and their hyperlinks comprising the World Wide Web (WWW)
can be represented as sets of nodes (vertices) and directed links
(edges). Such representations are referred to as complex networks
 or graphs~\cite{Newman2003}. Recently, much study has been devoted
to dynamics on complex networks in relation to, for example, the
spread of epidemics through  human travel
networks~\cite{Colizza2006}, the propagation of viruses through
the Internet~\cite{PWang2009}, information diffusion or consensus
of opinion in social acquaintance
networks~\cite{Sood2005,Baxter2008,SWSon2009}, and energy or
metabolite flux in ecological and metabolic
networks~\cite{DHKim2009}. In addition, applications of  network
dynamics have been explored as powerful predictive mechanisms for
elucidating community structure~\cite{Rosvall2008,YKim2010} and
node centrality~\cite{Masuda2009}. Foremost among these various
applications is Google's ranking algorithm~\cite{Brin1998}, {\it
PageRank} (PR),  a centrality or importance measure based on
random walks.

As originally conceived, PR rates web pages according to a link
analysis that takes into consideration only the topological
structure of the Web, and not the contents of its
pages~\cite{Brin1998,Borodin2001}, using a fundamental dynamic
process, random walks~\cite{HughesBook,JDNoh2004}. The algorithm
assumes that a random surfer (walker), starting from a random web
page, chooses the next page to which it will move by clicking at
random, with probability $d$, one of the hyperlinks in the current
page. This probability is represented by a so-called `damping
factor' $d$, where $d \in (0,1)$. Otherwise, with probability
$(1-d)$, the surfer jumps to any web page in the network. If a
page is a dangling end, meaning it has no outgoing hyperlinks, the
random surfer selects an arbitrary web page from a uniform
distribution and ``teleports'' to that page. In the case that many
random surfers exhibit the same surfing behavior, such that a
stationary state exists, the density of the surfers at each page
indicates the relative importance of each web page, {\it i.e.},
its PageRank.

Denoting the total number of pages as $N$,  the elements of the
transition matrix $\boldsymbol{P}$ are defined such that if page
$j$ has an outgoing link to $i$ (here we ignore multiple links),
$p_{ij}= 1/k^{\rm out}_j$, where $k^{\rm out}_j$ is the outgoing
degree of page $j$, $p_{ij}=1/N$ if page $j$ is a dangling end
with no outgoing degree ($k^{\rm out}_j=0$), and $p_{ij}=0$,
otherwise. We can describe the evolution of the PR (column) vector
$\boldsymbol{\pi}$ by the equation
\begin{equation}
\boldsymbol{\pi}(t) = d \boldsymbol{P} \boldsymbol{\pi}(t-1) +
\frac{(1-d)}{N} \boldsymbol{1}, \label{eq:1}
\end{equation}
where the column vector $\boldsymbol{1}=[1,\ldots,1]^T$,
$\pi_i(t)$ is the probability to be on page $i$ at time $t$, and
$d$ is the damping factor. If we write Eq.~(\ref{eq:1})
element-wise,
\begin{equation}
\pi_i(t) = d \sum_{j} p_{ij} \pi_j(t-1) + \frac{(1-d)}{N}.
\label{eq:PageRank}
\end{equation}
For $t \rightarrow \infty$, $\boldsymbol{\pi}(t)$ converges to the
stationary state which is given by the solution of the linear
system $( \boldsymbol{I} - d \boldsymbol{P}) \boldsymbol{\pi} =
\frac{(1-d)}{N} \boldsymbol{1}$. Alternatively, 
PR can be described by $\boldsymbol{\pi}(t) = \boldsymbol{G}
\boldsymbol{\pi}(t-1)$ with the Google matrix $\boldsymbol{G}$
defined as
\begin{equation}
\boldsymbol{G} = d \boldsymbol{P} + \frac{(1-d)}{N}
\boldsymbol{E},
\end{equation}
where the matrix $\boldsymbol{E}=\boldsymbol{1} \boldsymbol{1}^T$
is $1$ for all elements. The Google matrix is a Markov matrix. For
$d < 1$ it is (left) stochastic, aperiodic, and irreducible. The
PR vector $\boldsymbol{\pi}$ is the principal eigenvector, {\it
i.e.}, the eigenvector corresponding to the largest eigenvalue 1,
of the system $\boldsymbol{G} \boldsymbol{\pi} =\boldsymbol{\pi}$.
If $\pi_i > \pi_j$, then page $i$ ranks above page $j$.

Even though the damping factor $d$ is introduced mainly to prevent
the Google matrix from being reducible, its effect on the PR is
substantial. 
When $d$ goes to 0, the random teleport process dominates. The
result is a uniform state where all $\pi_i=1/N$. On the other
hand, as $d$ approaches 1, the transition process dominates, which
might suggest that the resulting PR reflects the network structure
more accurately. However, the resulting PR is not actually a good
indicator for finding important nodes. As Boldi {\it et al}. point
out, in the limit $d \rightarrow 1$, random walkers trivially
concentrate in the {\em rank-sinks}~\cite{Boldi2005,Boldi2006},
nodes (or groups of nodes) which have incoming paths but no
outgoing paths. Many important nodes will therefore have zero PR
in this limit.

As a result, choosing the damping factor close to 1 does not
provide a PR that indicates important nodes. Moreover when the
value of $d$ changes, not only can the PR values change
significantly~\cite{Pretto2002}, but also the relative rankings
can be radically altered~\cite{Langville2003,Lempel2004}, a
process called {\em rank-reversal}. When we consider that a
top-of-list Google ranking is deeply related to {\it e.g.} the
success of businesses and sales, rank-reversal stemming from
damping factor changes is of more than purely theoretical
relevance. Here we investigate rank-reversal dependence on the
damping factor and discuss its origin in directed networks.

\section{Strongly connected component decomposition}

\begin{figure}
\centering
\includegraphics[width=0.7\textwidth]{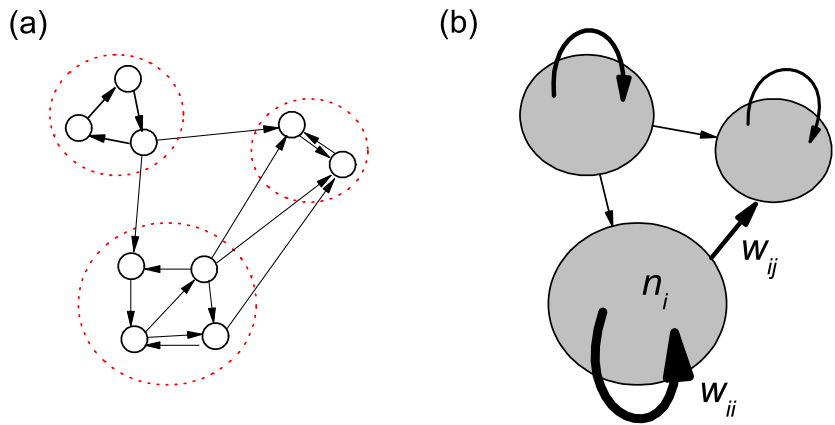}
\caption{ (Color online) SCC diagram. (a) A directed network can
be decomposed into several SCCs. Each red-dotted circle in (a)
delineates a SCC. (b) A directed network can also be abstracted
into a SCC diagram through coarse-graining. This abstracted SCC
diagram is an acyclic weighted network, containing self-links
(denoted by $w_{ii}$) and size-heterogeneous nodes ($n_i$). }
\label{fig:SCC_diagram}
\end{figure}

Because hyperlinks are directed, a random surfer may be able to
hop from page $A$ to page $B$ (possibly through several
hyperlinks), but may find that the return journey (from B to A) is
impossible. If this is the case, pages (nodes) $A$ and $B$ are
{\em weakly connected}. If, however, the surfer can return to $A$
from $B$ along a path of hyperlinks, $A$ and $B$ are considered to
be {\em strongly connected}. A {\em strongly connected component}
(SCC) is a maximal subnetwork of a directed network, such that
every pair of nodes in it is strongly connected; likewise, a
maximal weakly connected subnetwork is a {\em weakly connected
component} (WCC). A directed network can therefore be naturally
decomposed into one or more SCCs without ambiguity. Tarjan's
algorithm~\cite{Tarjan1972} efficiently finds the SCCs of a
directed network by performing a single depth-first search. If we
map each SCC to a virtual node (coarse-graining), a directed
network is abstracted as an acyclic weighted network, including
self-links, where the size of each virtual node represents the
number of nodes contained in its SCC (See
Fig.~\ref{fig:SCC_diagram}).

The largest SCC is called the giant strongly connected component
(GSCC), and every SCC (including single nodes) having paths {\it
to} the GSCC belongs to an incoming component. Conversely every
SCC having paths {\it from} the GSCC is called an outgoing
component. If a random walker starts from any node in the network
of Fig.~\ref{fig:SCC_diagram}(a) and follows only the directed
links without teleporting, the random walker will ultimately be
trapped in the top right SCC. This type of ``recurrent'' SCC is
called a {\em rank-sink}.

To prevent this situation, the damping factor is used.
Nonetheless, depending on one's choice of $d$, nodes can obtain
different ranking. Rank stability refers to how close relative
rankings are for different damping factors. 

\section{Correlation Coefficients}

To quantify rank-stability we use three different correlation
measures. The first is the well-known Pearson correlation
coefficient, which is defined for two paired sequences $(X_i,
Y_i)$ as follows,
\begin{equation}
r = \frac{\sum_{i=1}^{N} \left( X_i - \bar{X} \right) \left( Y_i -
\bar{Y} \right)}{\sqrt{\sum_{i=1}^{N} \left( X_i - \bar{X}
\right)^2}\sqrt{\sum_{i=1}^{N} \left( Y_i - \bar{Y} \right)^2}},
\end{equation}
where $\bar{X}$, $\bar{Y}$ are the sample mean of each variable.
Even though the Pearson correlation coefficient is widely used
across disciplines, owing to its invariance under scaling and
relocation of the mean, it is 
not robust~\cite{Wilcox2005}, as it strongly depends on outliers
in heavy-tailed data~\cite{Raschke2010}. This is a serious
problem, as many complex networks show heterogeneous degree
distributions and the PR vectors are also heavy-tailed. Therefore
we also use two other measures -- the Spearman and Kendall rank
correlation coefficients, which are known to be
robust~\cite{Raschke2010}. These also deal specifically with
relative ranks than absolute PR values, which is more relevant for
searches.

Defining $x_i$ and $y_i$ as the rank of $X_i$ and $Y_i$,
respectively, the Spearman correlation $r_S$ between $X$ and $Y$
is just the Pearson correlation coefficient between the ranks $x$
and $y$. It is defined as
\begin{equation}
r_S = \frac{\sum_{i=1}^{N} \left( x_i - \bar{x} \right) \left( y_i
- \bar{y} \right)}{\sqrt{\sum_{i=1}^{N} \left( x_i - \bar{x}
\right)^2}\sqrt{\sum_{i=1}^{N} \left( y_i - \bar{y} \right)^2}} =
1 - \frac{6 \sum_{i=1}^{N} (x_i - y_i)^2 }{N(N^2-1)}.
\end{equation}
The Spearman correlation reflects the monotonic relatedness of two
variables. The Kendall rank correlation coefficient, on the other
hand, counts the difference between the number of concordant pairs
and the number of discordant pairs, according to
\begin{equation}
\tau = \frac{ \sum_{i=1}^{N} \sum_{j=1}^{N} {\rm sgn} \left[ ( x_i
- x_j )( y_i - y_j )\right] }{N(N-1)},
\end{equation}
where ${\rm sgn}(x)$ is the sign function, which returns 1 if
$x>0$; $-1$ if $x<0$; and 0 for $x=0$. Here $( x_i - x_j )( y_i -
y_j )>0$ means concordant, and negative means discordant. The
Kendall rank correlation coefficient is typically used to quantify
{\em rank-stability} and {\em rank-similarity}~\cite{Lempel2004}.

\section{Dataset}

Here we use the Stanford Web data, collected in
2002~\cite{StanfordData,Leskovec2008}, from Stanford University's
Internet domain (stanford.edu). The data contains 281,903 web
pages and 2,312,497 hyperlinks. Even though this is, itself, a
subsample of the whole WWW, it is an accurate representation of a
part of the Web since it was not gathered by a crawler, but
instead contains all web pages in a single
domain~\cite{SWSon2012}. It exhibits the topological
characteristics of the WWW~\cite{Leskovec2008,Broder2000}. (See
the detailed topological properties of the Stanford network data
in the Appendix.)

\begin{table}[h]
\caption{\label{data} Summary of the Stanford Web data.}
\begin{tabular}{lccccc} \br
Number of & Number of & Average & Number of & Size of & Degree-degree \\
nodes & links & degree & SCCs & the giant SCC & auto-correlation \\
\mr
281,903 & 2,312,497 & 8.20 & 29,914 & 150,532 (0.534) & 0.047 \\
\br
\end{tabular}
\end{table}

\section{Effects of damping factor on PageRank}

\begin{figure}
\centering
\includegraphics[width=0.47\textwidth]{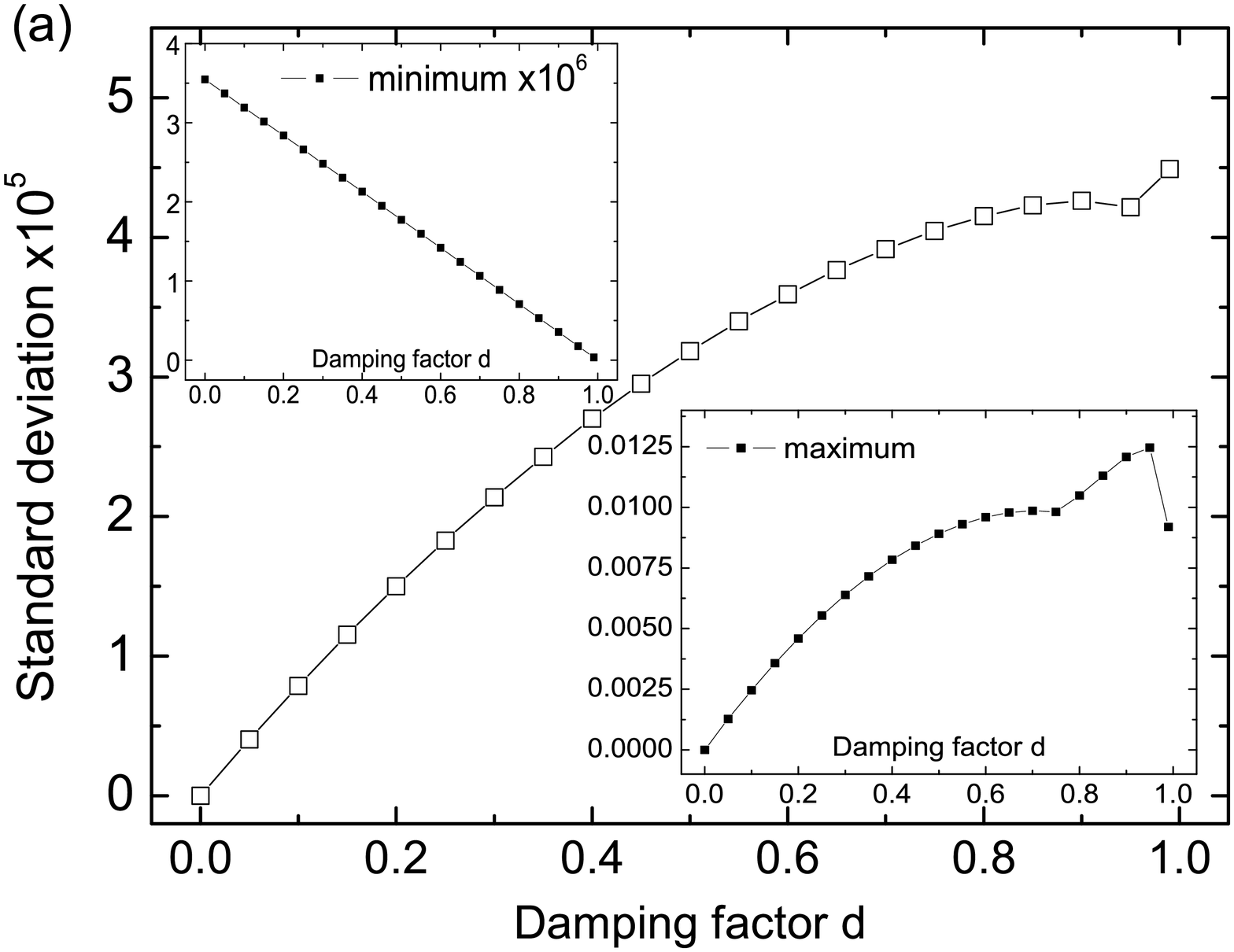}
\includegraphics[width=0.49\textwidth]{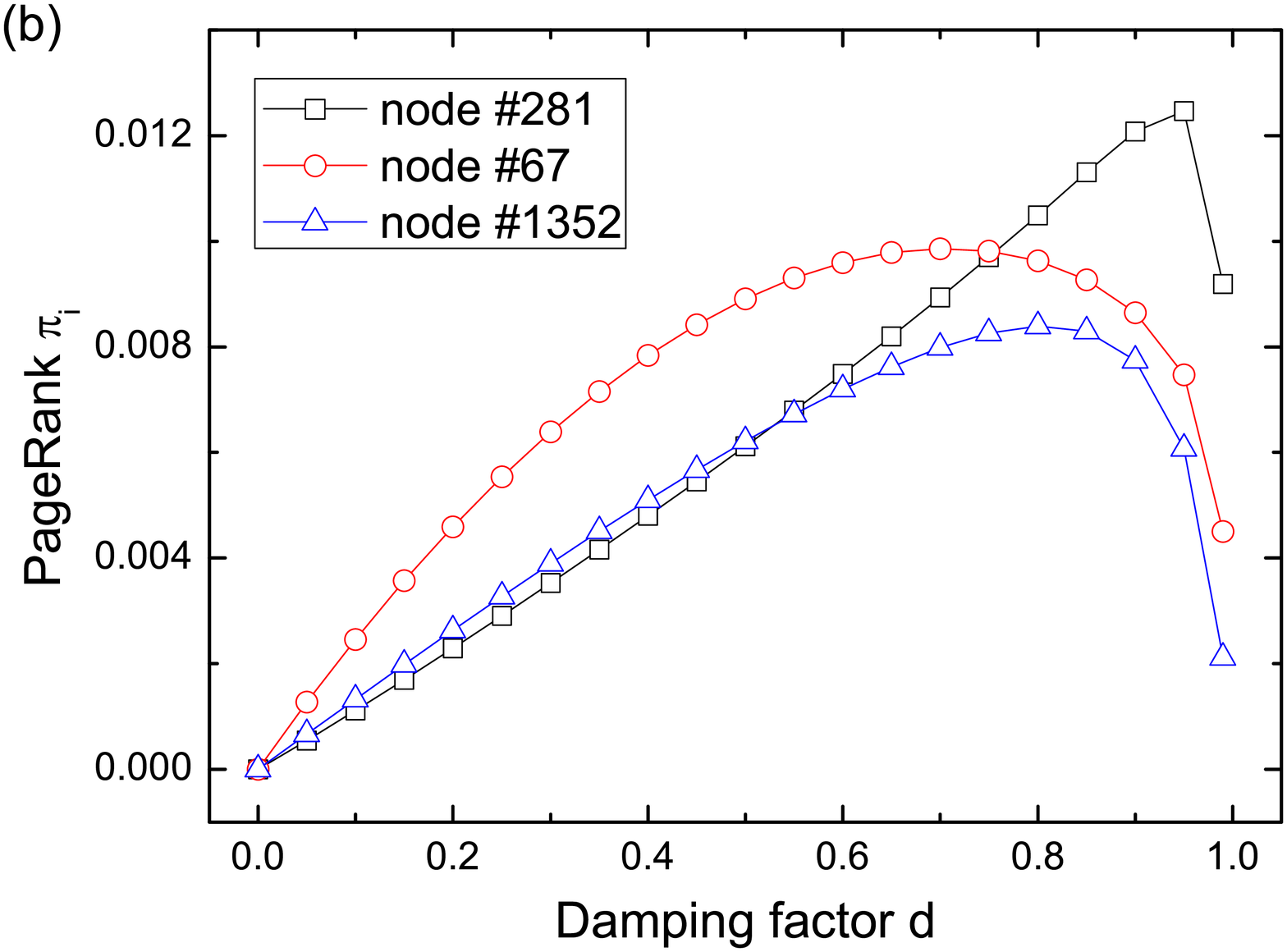}
\caption{ (Color online) (a) Standard deviation of the PR of the
Stanford network, along with its minimum and maximum. The standard
deviation of the PR gradually increases as the damping factor
increases. In the insets: the minimum of the PR follows a trivial
linear relation, while the maximum is nontrivially correlated with
the value of the damping factor. (b) PR values of the three nodes
having the highest PR are traced as the value of the damping
factor is changed. Rank reversals occur around $d=0.55$ and
$d=0.75$. } \label{fig:PageRank_dist}
\end{figure}

\begin{figure}[b]
\centering
\includegraphics[width=0.50\textwidth]{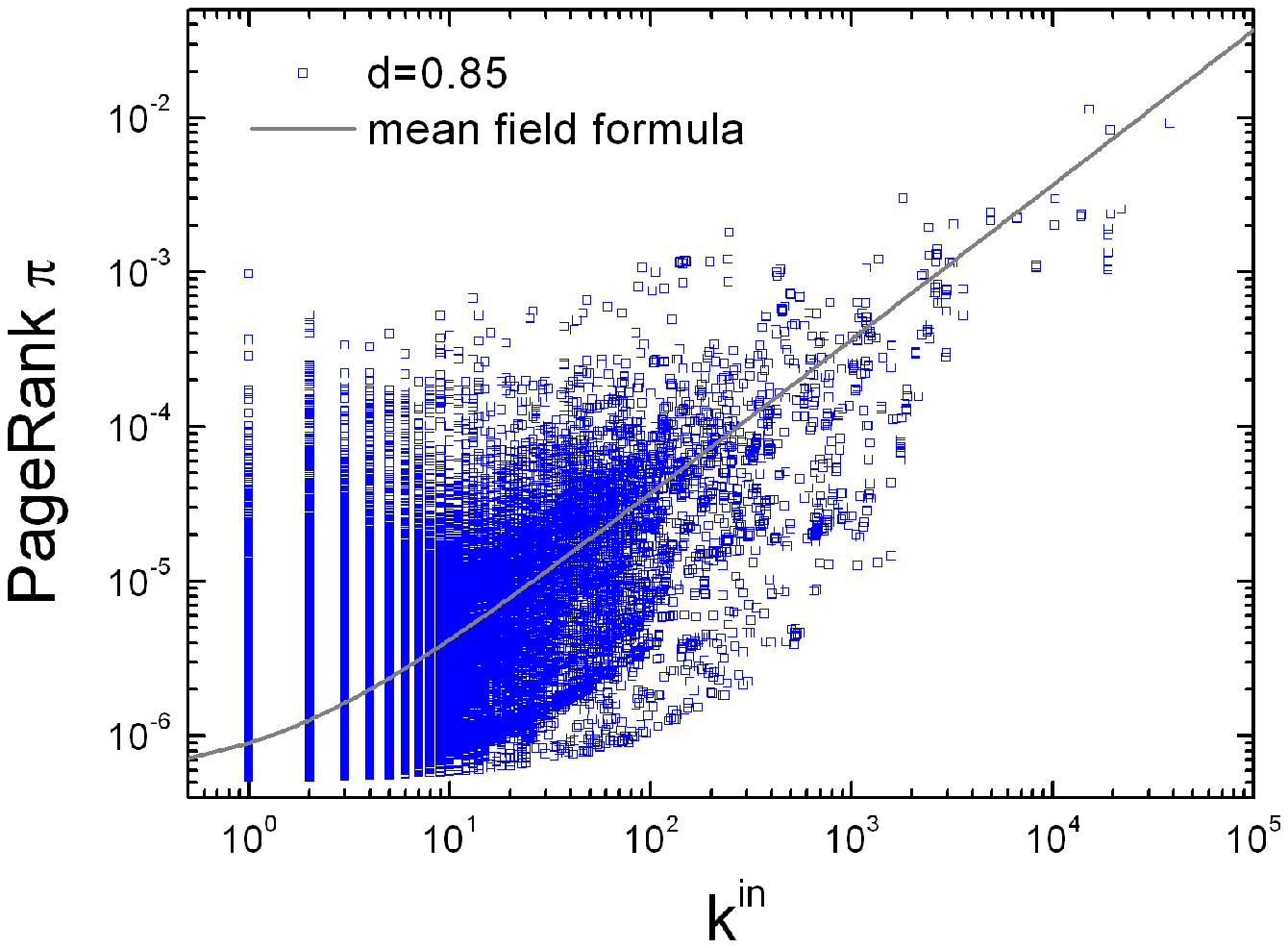}
\includegraphics[width=0.475\textwidth]{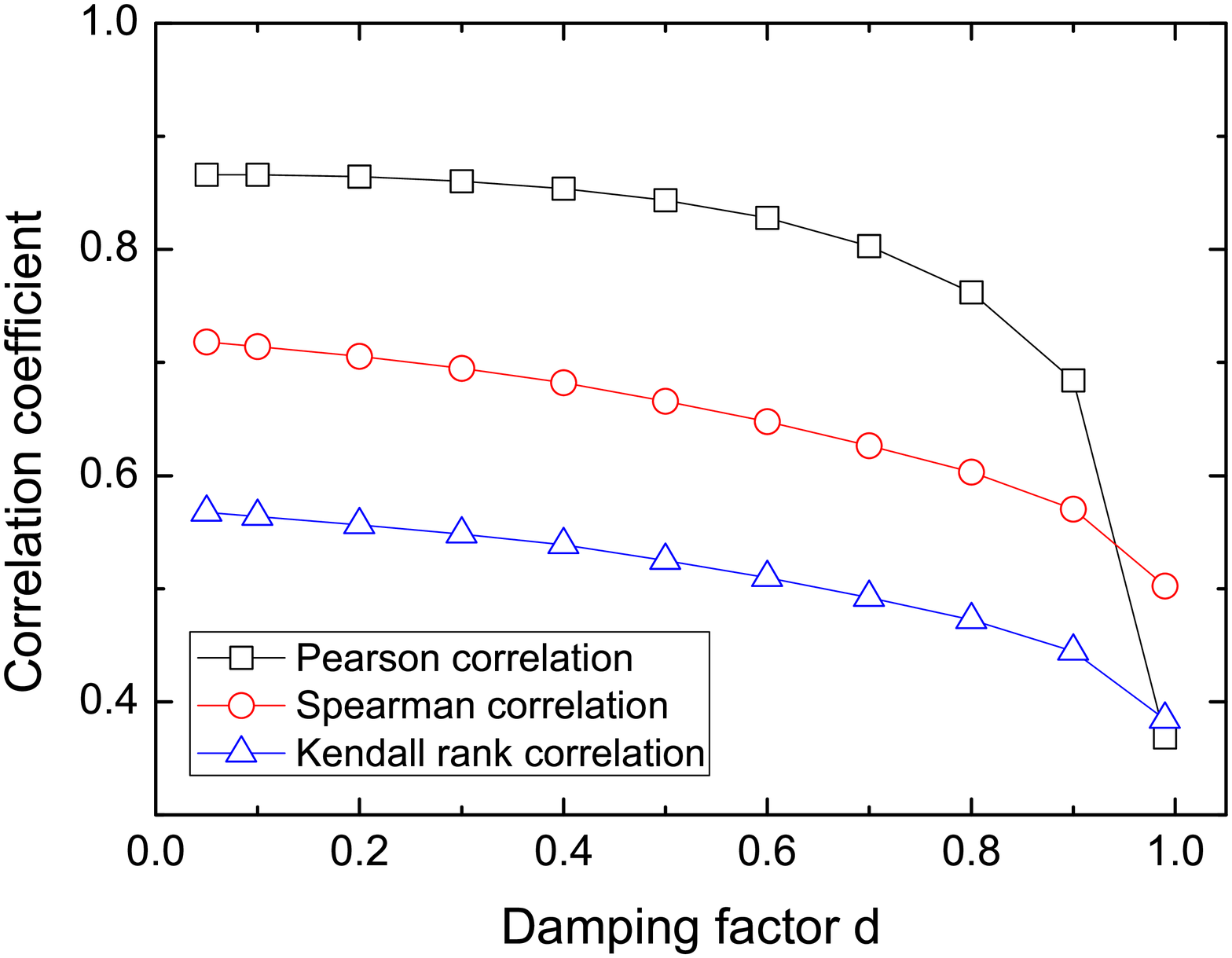}
\caption{ (Color online) The relationship between incoming degree
and PR. The left panel clearly shows a positive correlation
between the incoming degree and PR for $d_0=0.85$. It agrees well
with the mean field result of Ref.~\cite{Fortunato2007}, indicated
by the solid line. The right-side panel shows the three different
correlation coefficients between incoming degree and PR at
different values of $d$. The correlation increases as the damping
factor decreases. } \label{fig:kin_PR}
\end{figure}

We first examine the responses of PR to changes in the value of
the damping factor $d$, as quantified by its standard deviation,
minimum, and maximum. As can be seen from
Fig.~\ref{fig:PageRank_dist}(a), the standard deviation of the PR
gradually increases as the damping factor changes from zero to
one. When the damping factor is zero, every PR is $1/N$ since in
this regime only teleportation occurs. Therefore, both the minimum
and maximum of the PR are $1/N$. The minimum of the PR (in
Fig.~\ref{fig:PageRank_dist}(a), the left inset) decreases
linearly as the damping factor increases since nodes with no
incoming degree attain the minimum PR, $(1-d)/N$. However, the
maximum of the PR is not trivial (in
Fig.~\ref{fig:PageRank_dist}(a), the right inset). It seems to
depend on the topology of the network. To understand the behavior
of the maximum PR we follow the behavior of the three nodes of
highest PR, as shown in Fig.~\ref{fig:PageRank_dist}(b). For any
value of $d$, the maximum of PR is always associated to one of
these nodes. These three nodes, situated in the giant SCC, show
{\em rank-reversal} as the damping factor changes, thus accounting
for the anomalous behavior of the maximum.

A correlation between incoming degree and PR has been reported by
Fortunato {\it et al}.~\cite{Fortunato2007} and is confirmed in
Fig.~\ref{fig:kin_PR}. The correlation coefficients between
incoming degree and PR are $(r, r_S, \tau)=(0.730, 0.589, 0.460)$
at $d=0.85$. As shown in the left-side panel, nodes with large
incoming degree exhibit relatively narrow fluctuations in PR,
while nodes of small degree show larger fluctuations. As the
damping factor increases, the fluctuations of small degree nodes
become broader and seem to induce smaller correlation coefficients
at higher damping factors (See Fig.~\ref{fig:kin_PR}, the
right-side panel). The correlation between PR and incoming degree,
on the other hand, becomes higher as the random teleporting
process increases. In that case, random walkers follow fewer steps
in the network before teleporting and therefore sense only
incoming degree, rather than higher-order link-link correlations.

\section{Rank-reversal under damping factor perturbations}

\begin{figure}[bh]
\centering
\includegraphics[width=0.45\textwidth]{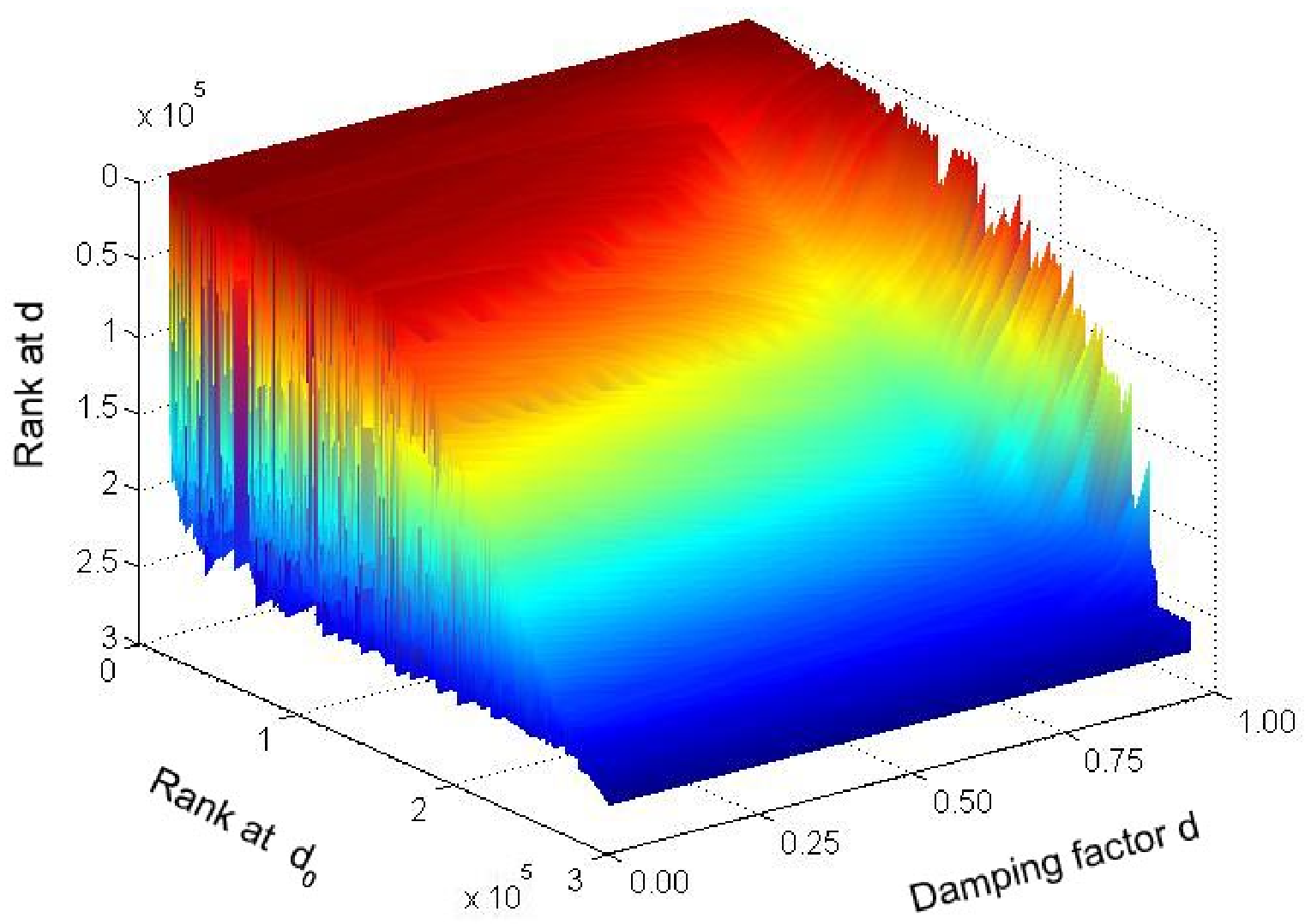}
\includegraphics[width=0.45\textwidth]{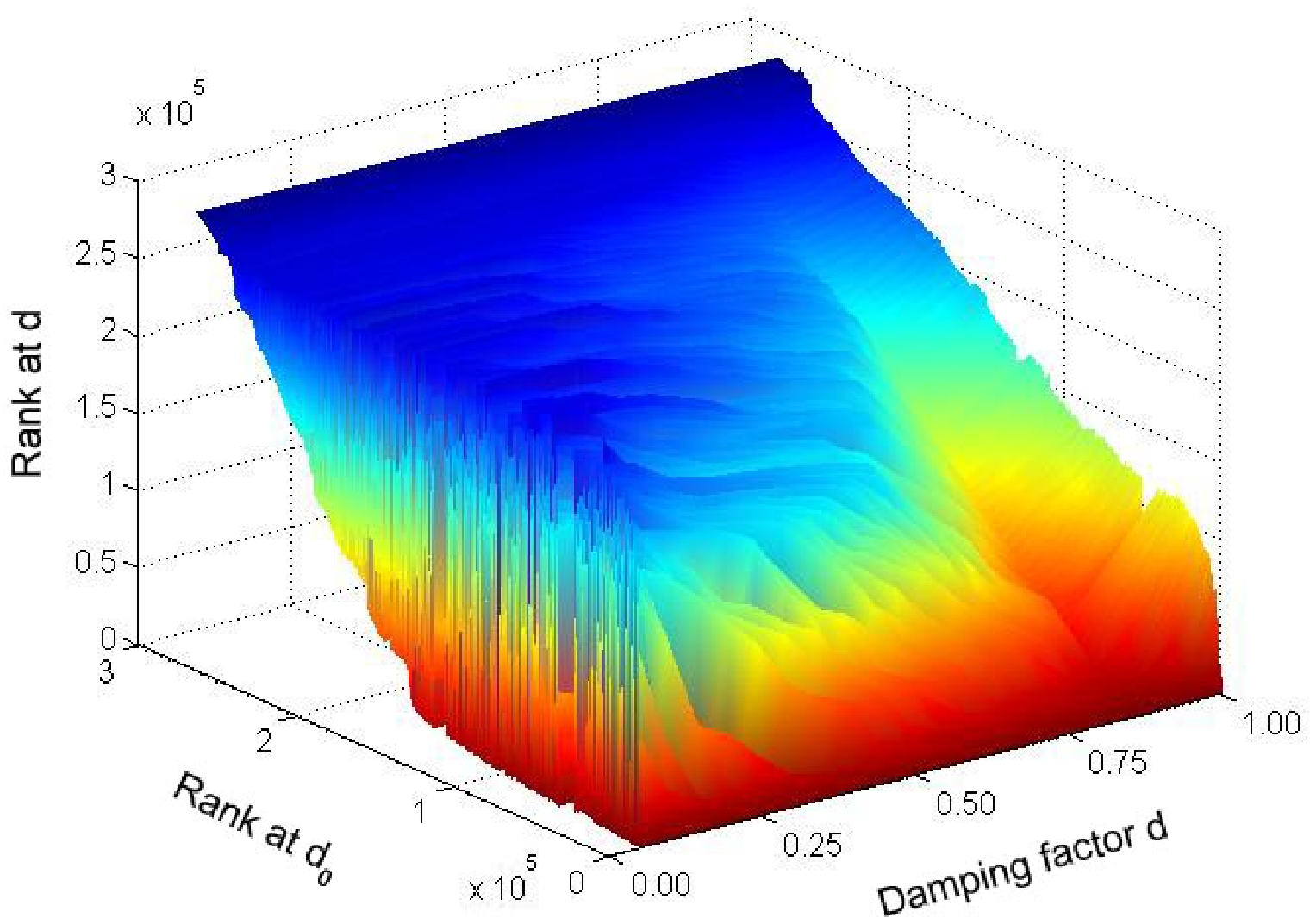}
\includegraphics[width=0.52\textwidth]{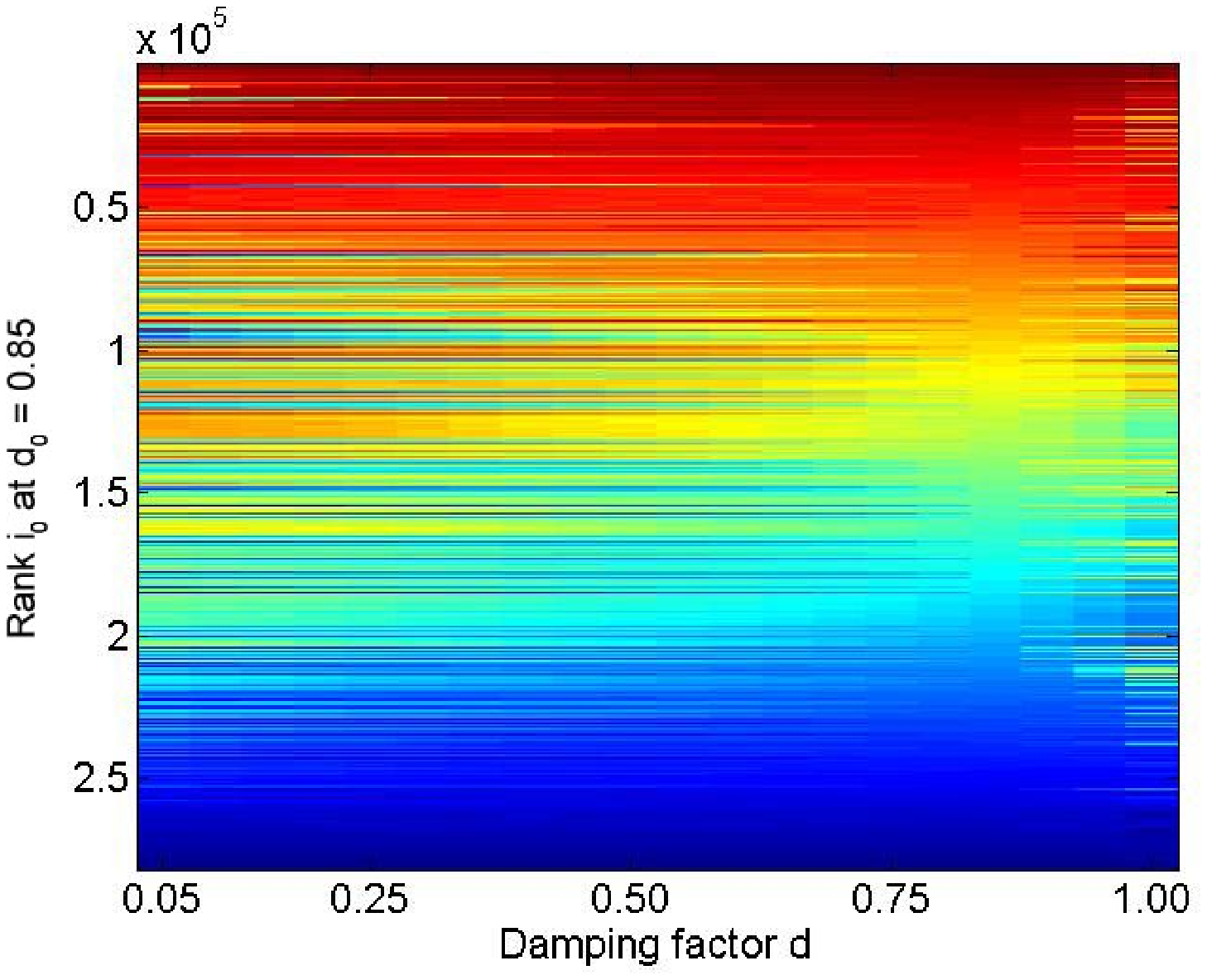}
\caption{ (Color online) Rank changes depending on the damping
factor $d$. At $d_0=0.85$, nodes are sorted in descending order
and are assigned an index $i_0$. Bottom figure: Color corresponds
to the rank at the value $d$ given on the horizontal axis. Red
represents the highest ranking and blue lowest. When the damping
factor changes from 0.05 to 0.99, we observe how the rank changes.
The top two plots give the rank changes in 3D (rank for $d$ is the
z-axis). The rightmost plot is simply the leftmost plot shown from
behind. } \label{fig:Color_Rank}
\end{figure}

Figure~\ref{fig:Color_Rank} shows the overall change of the ranks
for 281,903 nodes in response to damping factor changes away from
$d_0=0.85$. We sort the nodes in descending order of PR at
$d_0=0.85$ and assign to each node both a label $i_0$ as well as a
corresponding color. Red represents the highest rank, blue the
lowest. When the damping factor changes, we observe how the rank
of each node changes. The top two plots give the rank changes in
3D (rank is the z-axis). The rightmost plot is the same as the
leftmost plot but shown from behind. It demonstrates that rank
changes occur often near the extremes of $d=0.05$ and $d=0.99$.
Rank reverses near $d=0.85$ are quite apparent, and clear rank
changes in this regime are observed over all nodes of the network.
It seems that the middle-rank nodes show more rank change than do
the top- and bottom-rank nodes near $d_0=0.85$.

\begin{figure}[b]
\centering
\includegraphics[width=0.45\textwidth]{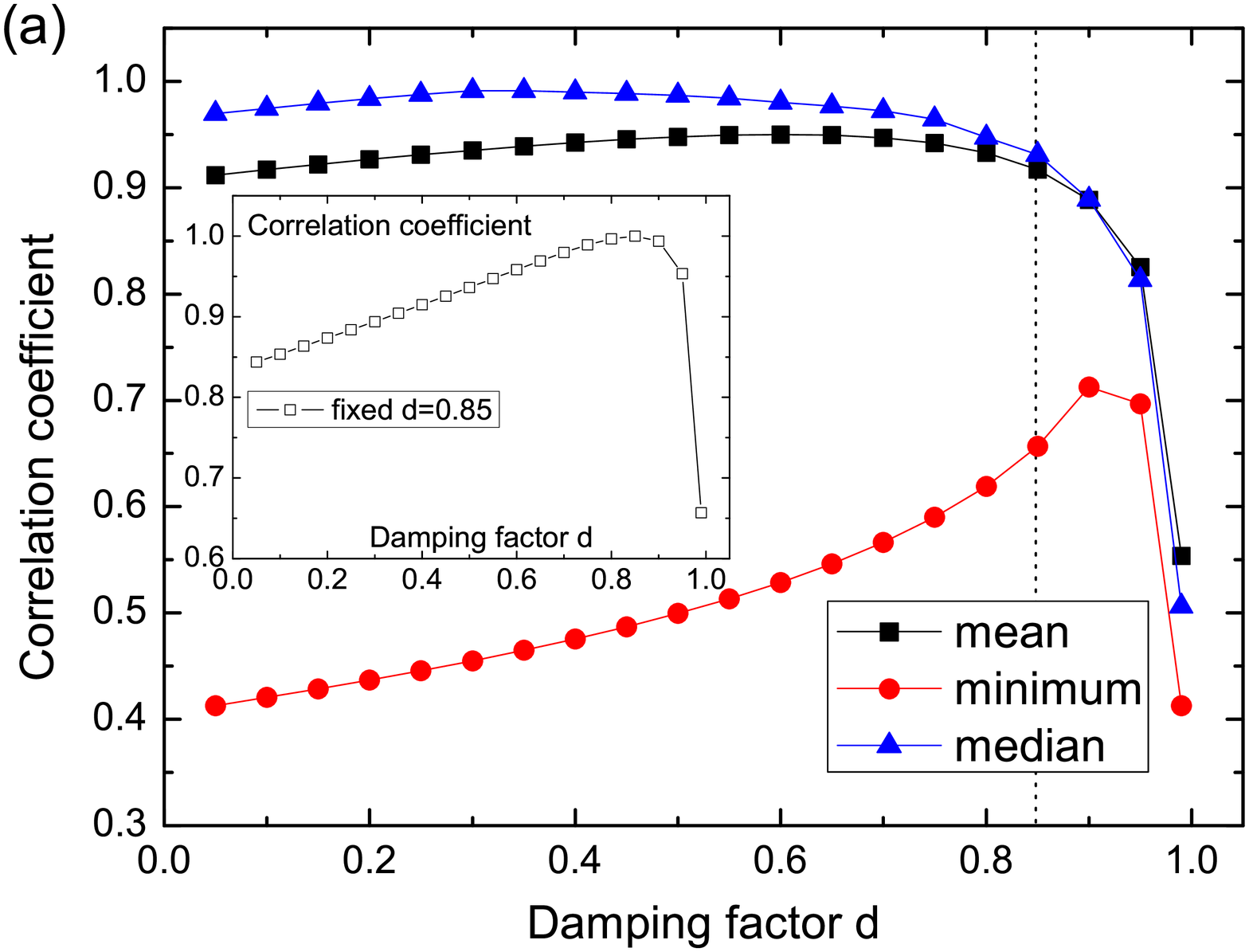}
\includegraphics[width=0.45\textwidth]{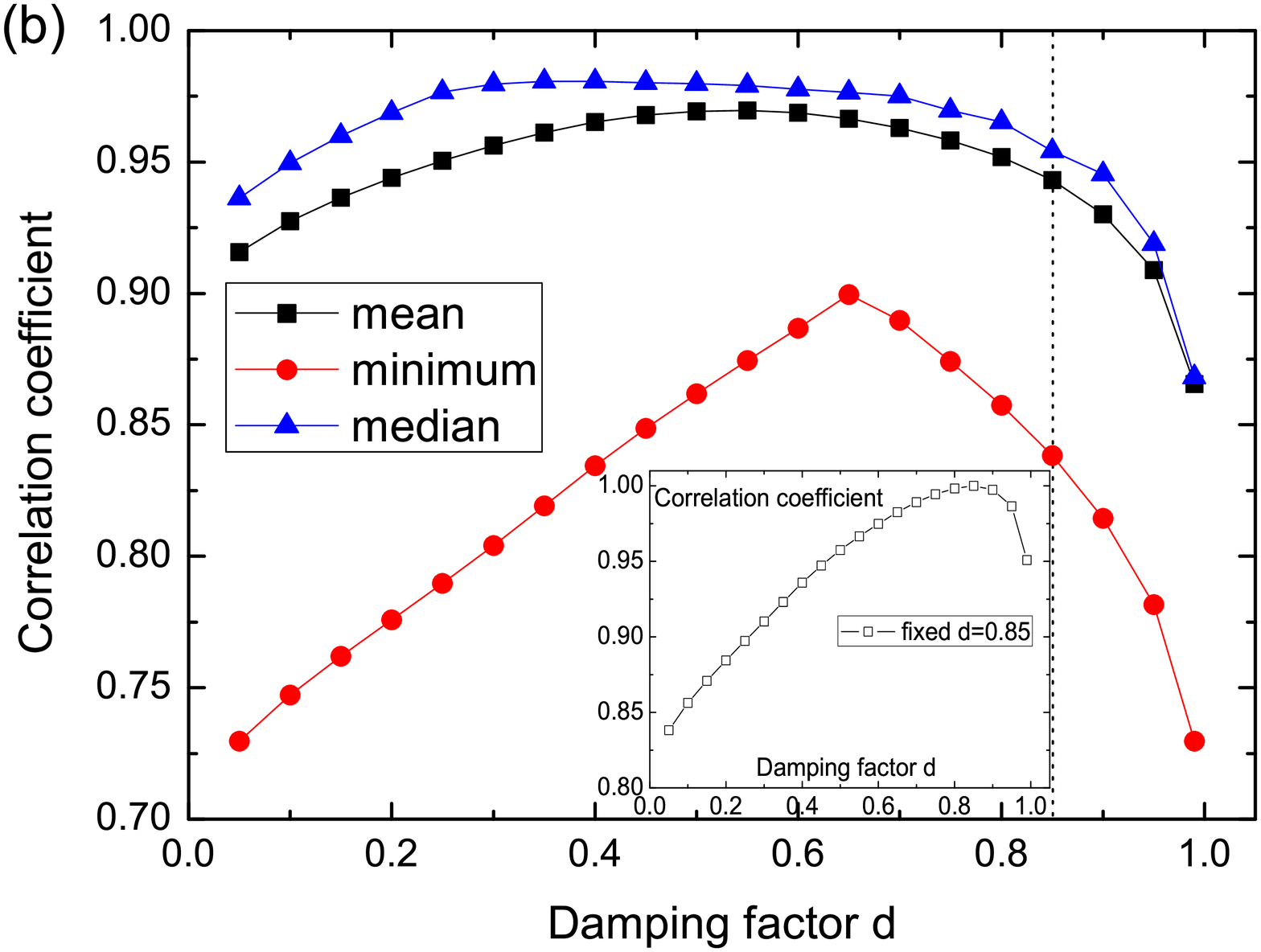}
\includegraphics[width=0.45\textwidth]{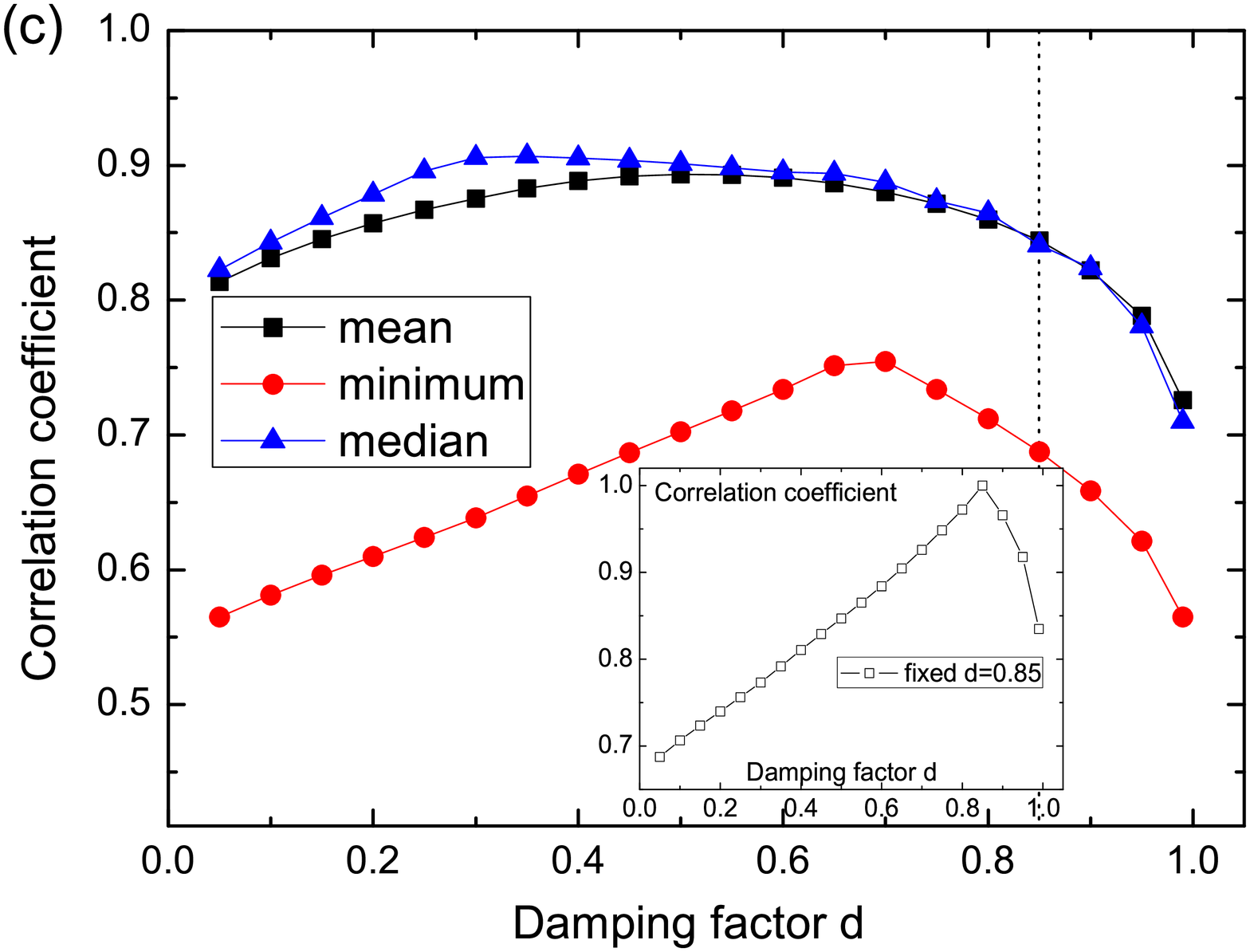}
\includegraphics[width=0.46\textwidth]{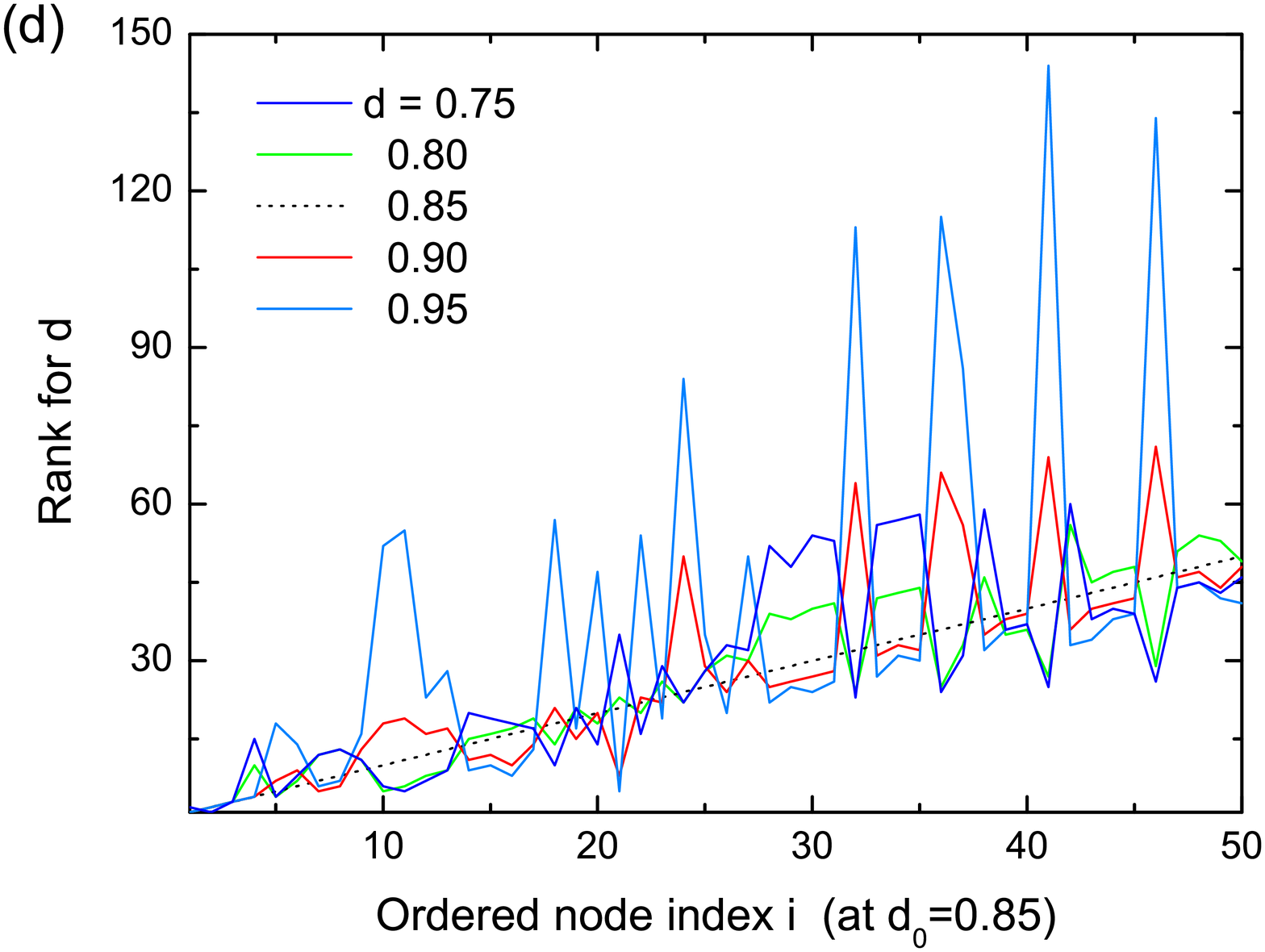}
\caption{ (Color online) The minimum, mean, and median of
correlation coefficients of Pearson correlation (a), Spearman
correlation (b), and Kendall rank correlation (c) between the PRs
for different values of the damping factor. The inset of each
figure illustrates the correlation coefficient between responses
of PR at $d=0.85$ and its responses at the other damping factor
values. (d) The 50 top-ranked nodes at $d=0.85$. While the damping
factor changes by only 0.1 from 0.85, even the top 50 nodes are
subject to relative rank fluctuations of up to three times their
original rank. } \label{fig:Pearson}
\end{figure}

In order to understand the sensitivity of PR to deviations in the
value of the damping factor, we measure the correlation
coefficients between two PR vectors at different $d$ values. For
20 values of the damping factor, $d=0.05, 0.1, \cdots, 0.95$ which
are equally spaced except for the last one $d=0.99$, we compute
the PRs and calculate correlation coefficients $C_{d d'}$ of PR
vectors for the 190 distinct damping factor pairs $(d,d')$. For
every $d$, we look at the minimum correlation coefficient
$C_{d,\rm{min}} = \min_{d'} C_{d d'}$ at a given $d$, along with
the mean and median of the distribution for the three correlation
definitions. For comparison, we also compute these correlations
for $d_0=0.85$ (the value used by Google), and display $C_{d d_0}$
in the inset of each plot for different correlation measures.

Figures~\ref{fig:Pearson}(a)-(c) show the behaviors of the mean,
median, and minimum of the three correlation coefficients for PR
at a given damping factor $d$ compared to the other 19 values of
the damping factor. The Pearson correlation in panel (a) shows
that a peak of the minimum correlation coefficient line occurs
around $d=0.90$ and decreases substantially for larger $d$. On the
other hand, the other correlation coefficients, which are
measurements of relative rank, place the peak of the minimum
around 0.65. The relative rank given by PR is more stable around
$d=0.65$ than it is around $d_0=0.85$ in this Stanford Web
network, when we consider minimal rank-reversal. We do not know to
what extent this result generalizes to other networks.

The panel insets relay the correlation coefficients between the
PRs at $d_0=0.85$ and the other values of $d$. Observing these
quantities, we can determine how much the PR changes when the
damping factor increases or decreases by 0.05 around $d_0=0.85$.
The Pearson correlation, in particular, suggests that the PR value
is very sensitive to changes in $d$ when the damping factor is
large. Rank changes in response to damping factor changes by 0.1
upward and downward are given in Fig.~\ref{fig:Pearson}(d) for the
50 top-ranked pages at $d_0 = 0.85$. Remarkably, even the
top-ranked nodes are subject to significant changes in rank
(relative rank may change by up to three times its original value)
when $d=0.95$. When $d$ increases 0.1 from $d_0=0.85$, the Kendall
rank correlation becomes 0.918, which means roughly $1.6 \times
10^9$ pairs (about 4\% of the total pairs) are rank-reversed.

\section{Rank-reversal in a single SCC}

\begin{figure}
\centering
\includegraphics[width=0.8\textwidth]{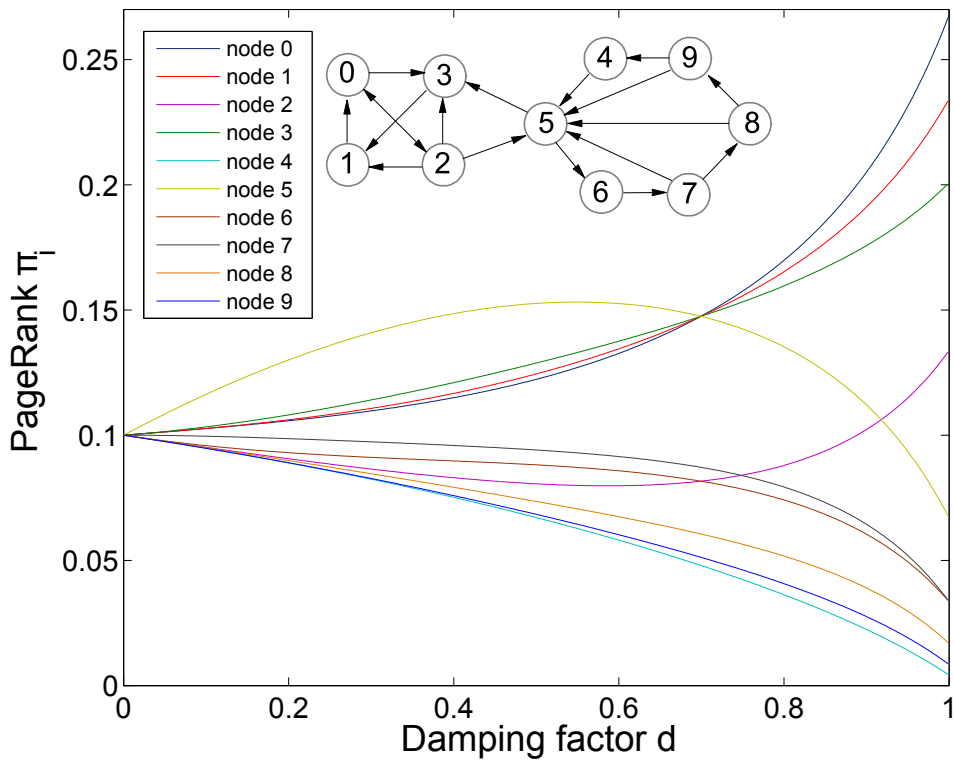}
\caption{ (Color online) Single SCC example of rank reversal.  A
directed network comprised of 10 strongly-connected nodes is,
itself, a SCC without any dangling nodes or sinks. As can be seen
in the figure, even this network undergoes rank reversal for high
values of the damping factor. } \label{fig:PageRankOneSCC}
\end{figure}

As Boldi {\it et al}. discuss in their
studies~\cite{Boldi2005,Boldi2006}, rank-reversals occur
frequently in directed networks as a result of dangling nodes and
rank-sinks. When the directed network contains a rank-sink, as the
damping factor approaches 1 PR becomes trivially concentrated in
the sink component(s). For this reason, choosing $d$ close to 1
does not give the best value of PR since many important nodes have
a null PR in the limit $d \rightarrow 1$. In fact a similar effect
occurs even when the directed network has no rank-sink, such as is
the case with a single SCC.

To further explore this phenomenon, we examine a simple example of
a directed network, comprised of 10 strongly-connected nodes (See
Fig.~\ref{fig:PageRankOneSCC}).  For this network, one can
explicitly write down the 10 PR equations of
Eq.~(\ref{eq:PageRank}):
\begin{eqnarray}
\nonumber \fl \qquad & \pi_0 = d \left( \pi_1 + \frac{\pi_2}{4}
\right) + \frac{1-d}{10}, \quad
& \pi_1 = d \left( \pi_3 + \frac{\pi_2}{4} \right) + \frac{1-d}{10}, \\
\nonumber \fl & \pi_2 = d~ \frac{\pi_0}{2} + \frac{1-d}{10}, &
\pi_3 = d \left( \frac{\pi_0}{2} + \frac{\pi_2}{4} +
\frac{\pi_5}{2} \right) +
\frac{1-d}{10}, \\
\fl & \pi_4 = d~ \frac{\pi_9}{2} + \frac{1-d}{10}, & \pi_5 = d
\left( \pi_4 + \frac{\pi_2}{4} + \frac{\pi_7}{2} + \frac{\pi_8}{2}
+
\frac{\pi_9}{2} \right) + \frac{1-d}{10}, \\
\nonumber \fl & \pi_6 = d~ \frac{\pi_5}{2} + \frac{1-d}{10}, &
\pi_7 = d~ \pi_6
+ \frac{1-d}{10}, \\
\nonumber \fl & \pi_8 = d~ \frac{\pi_7}{2} + \frac{1-d}{10}, &
\pi_9 = d~ \frac{\pi_8}{2} + \frac{1-d}{10}.
\end{eqnarray}
This system of 10 linear equations is solved and the results are
plotted in Fig.~\ref{fig:PageRankOneSCC} as a function of the
damping factor $d$. While the network has only one SCC, rank
reversals occur as the value of $d$ changes.

As one can see from this simple example, certain substructures of
a network, like `pockets' ({\it rank-pockets}), -- in this example
$\{0, 1, 2, 3\}$ and $\{4, 5, 6, 7, 8, 9\}$ -- can {\em
concentrate} the random walker inside causing rank-reversal. These
structures could be modular structures like
communities~\cite{Rosvall2008,YKim2010}. In the above example,
node 2 has outgoing degree 4. Among these outgoing links, however,
only one link points toward the outside of the module, making a
narrow channel. If we consider the extreme case that a node has
very large outgoing degree $k^{\rm out}$ and all but one of the
outgoing links of the node point toward the inside of the module
(or rank-pocket) and only one link points outside, the pocket
becomes a {\it rank-sink} in the limit $k^{\rm out} \rightarrow
\infty$. In other words, the pocket becomes a trapping structure
that is characterized by a vanishing `bottleneck'.

Uneven link density between modules (or other substructures) of a
network allows pockets to concentrate random walkers inside,
subsequently causing rank-reversal. These reversals call into
question again: which damping factor $d$ is the best choice? In
the example of Fig.~\ref{fig:PageRankOneSCC}, other centrality
measures such as degree and betweenness centralities support node
5 as being the most important, but PR only supports this ranking
when $d$ is less than about 0.7. The example, while simple,
indicates that the best choice for damping factor may depend on
the network structure, or on what features associated with a
node's position in a network one considers most important.

The structure of rank-pockets is very similar with that of `spam
farms'~\cite{Gyongyi2004, YeDu2007}. These are groups of web pages
that are intentionally interconnected to boost the PR of target
pages giving them higher rankings than they deserve by
``misleading'' the PR's link based algorithm. Since spam farms are
continuously optimized through trial and error for Google's `real'
damping factor and actual algorithm, filtering them is a
challenging and outstanding computer science
problem~\cite{YeDu2007, SHan2006}.

\section{Summary and concluding remark}

Given that the success of modern businesses or the ranking of
athletes~\cite{Radicchi2011}, scientists~\cite{Radicchi2009},
their papers~\cite{Chen2007}, or scientific
journals~\cite{Rosvall2008,Bergstrom2008} in which those papers
are published depends on Google's PageRank algorithm and its
resulting ratings, it is far from a purely academic venture to
understand rank-stability~\cite{ANg2001} and its dependence on
network structure~\cite{Ghoshal2011} and damping factor. Hence we
have investigated PageRank (PR) as a function of its damping
factor on a subset of pages from a single domain in the World Wide
Web and found that rank-reversal occurs frequently and over a
broader range of PR. We note that the Pearson correlation of PR
between two different damping factors rapidly drops as the damping
factor increases from the frequently-used value, 0.85.
Rank-reversal is also observed by measuring the Spearman
correlation and Kendall rank correlation. For this network the
most stable value of the damping factor, in terms of relative
rank, is about 0.65. This rank-reversal happens not only in
directed networks containing rank-sinks but also in a single
strongly connected component (SCC). This is due to the presence of
rank-pockets and bottlenecks. When the damping factor approaches
1, PR converges trivially such that many important nodes have tiny
PR possibly even within a single SCC. A better understanding
rank-reversals may be essential to optimizing the stability of PR,
to thwarting attempts to cheat such as spam farms, and ultimately
to determining which scientists will be cited, which products will
sell, and which businesses or other ventures will prosper.

\appendix

\section{Stanford Web network}
\begin{figure}[b]
\centering
\includegraphics[width=0.85\textwidth]{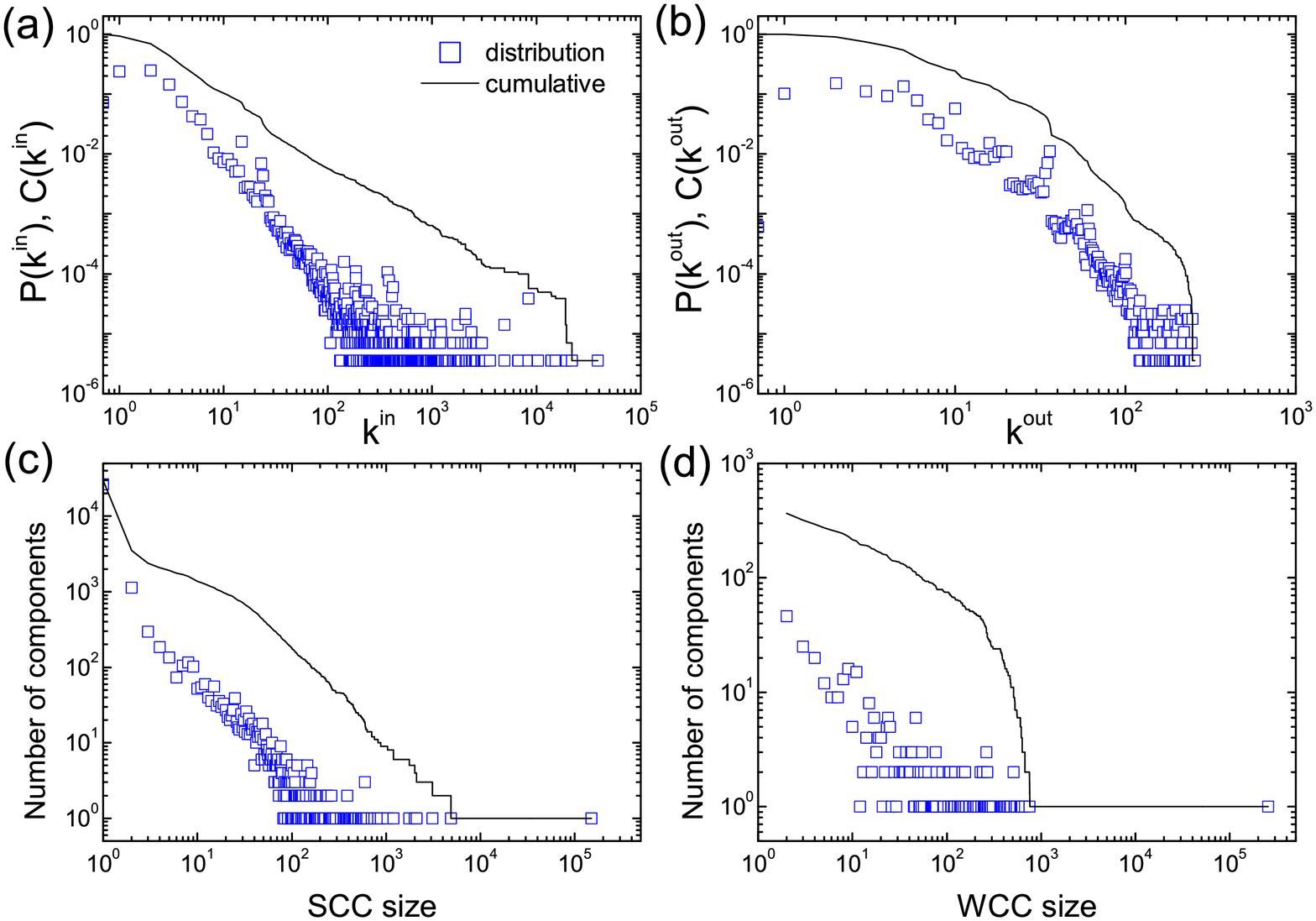}
\caption{ (Color online) (a) Incoming- and (b) outgoing-degree
distributions and (c) SCC and (d) WCC size distributions for the
Stanford Web network
. The solid black line in each plot is the complementary
cumulative distribution of the scatter plot. }
\label{fig:StanfordDegree} 
\end{figure}

\begin{figure}
\centering \vskip .05in
\includegraphics[width=0.85\textwidth]{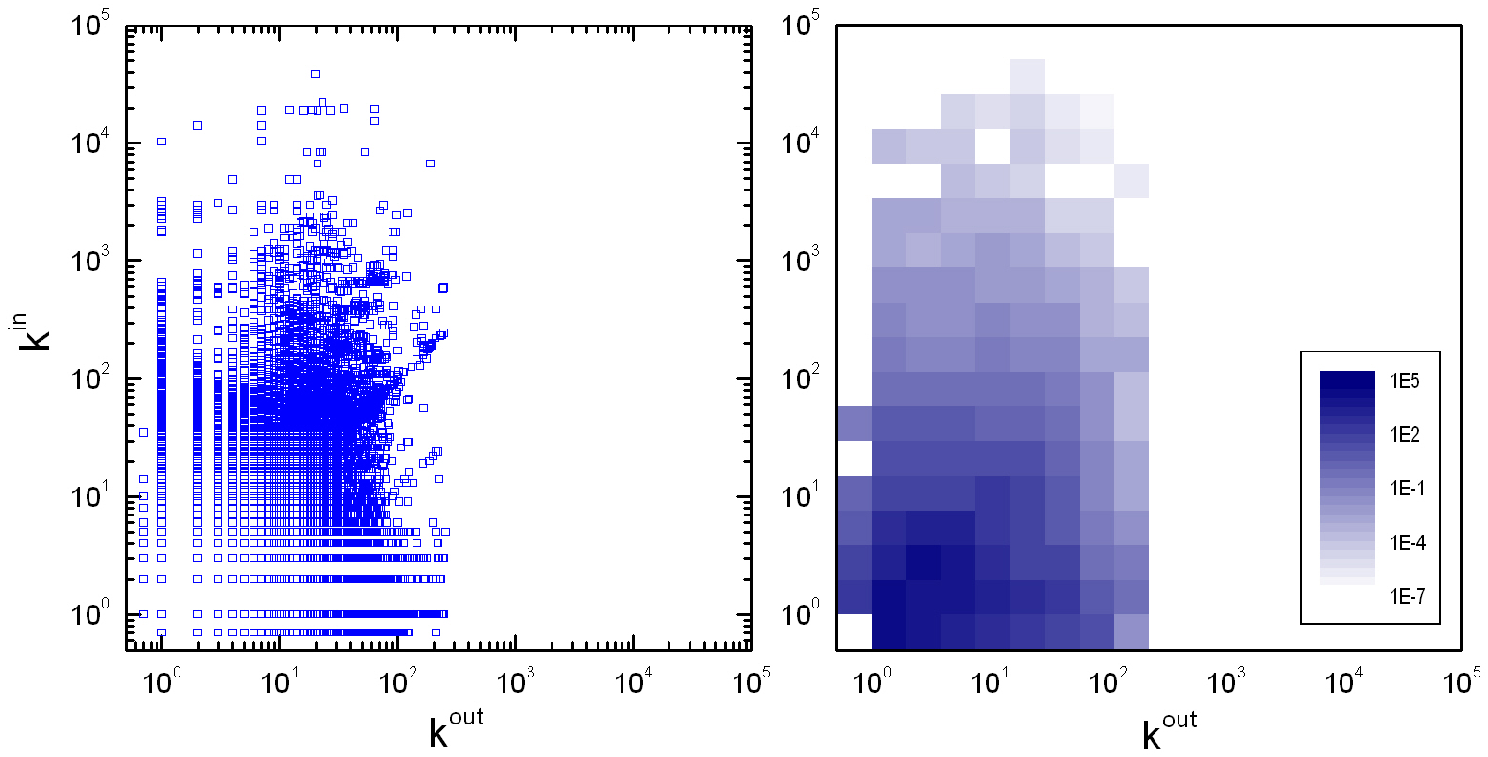}
\caption{ (Color online) Scatter plot of incoming and outgoing
degrees for each node (left) and it density plot (right). }
\label{fig:DegreeAutocorrelation}
\end{figure}

\begin{figure}
\centering
\includegraphics[width=0.8\textwidth]{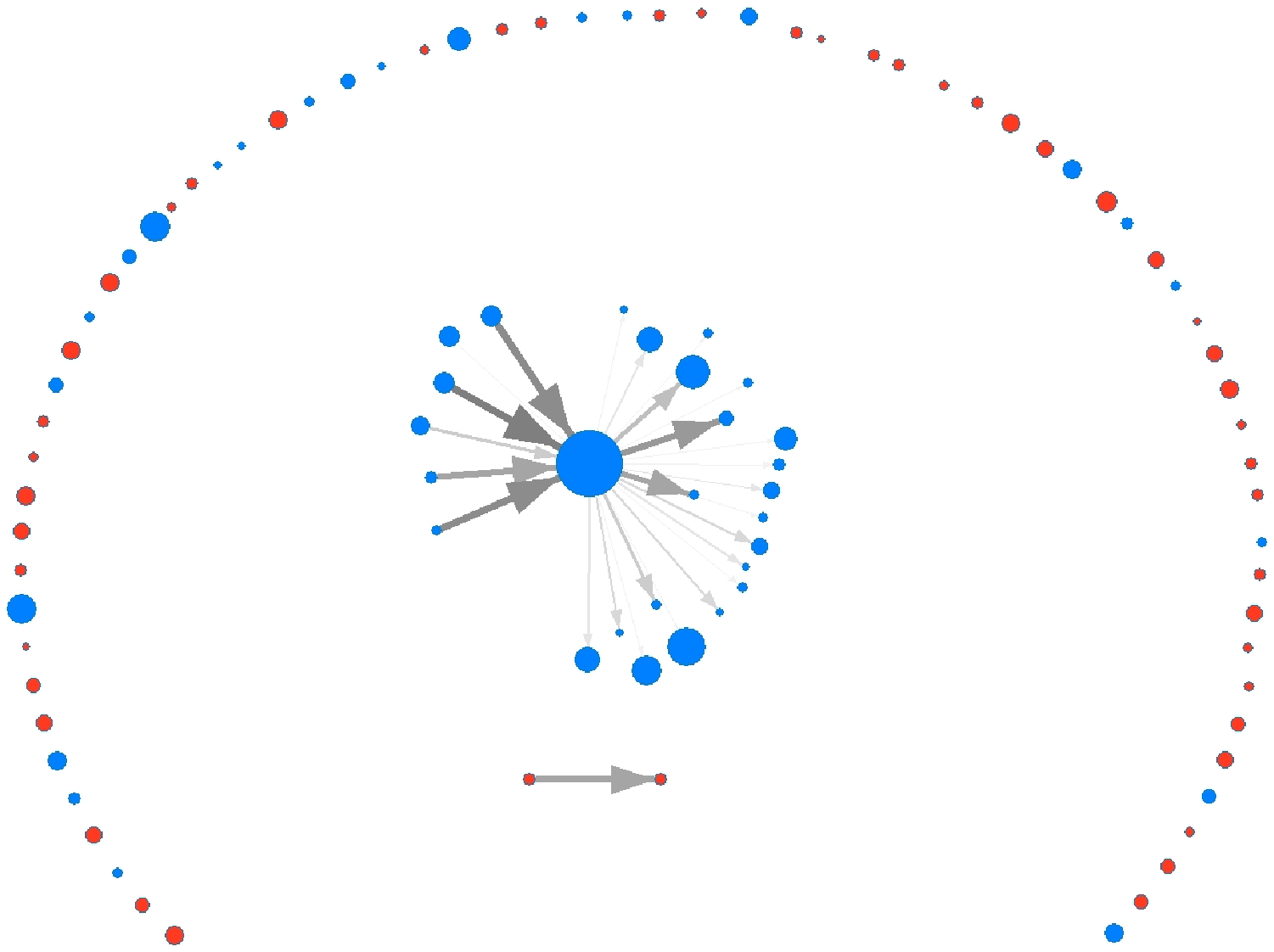}
\caption{ (Color online) SCC diagram for the Stanford Web network.
For visualization, only the 98 largest SCCs have been displayed.
Each circle corresponds to a SCC whose size is proportional to the
logarithm of the number of nodes in the SCC. The SCCs in the
largest WCC are colored blue, and the others red. The width and
gray-scaled color of the directed links reflect their weight.
Self-links are omitted. This SCC diagram shows a simple {\em
bow-tie} structure. } \label{fig:SCCdiagram}
\end{figure}

The Stanford Web network we study exhibits a broad incoming degree
distribution, that can be roughly characterized as a power-law
$P(k) \sim k^{\gamma}$ with $\gamma \approx 2$, but its outgoing
degree distribution is not easily classified (See
Fig.~\ref{fig:StanfordDegree}(a), (b)). Looking at the
degree-degree auto-correlation, the scatter plot of
Fig.~\ref{fig:DegreeAutocorrelation} shows no clear pattern in
scatter plot. Only the density plot indicates a vague positive
relationship between incoming and outgoing degrees in this
network. The Pearson, Spearman, and Kendall correlations between
incoming and outgoing degrees are 0.047, 0.258, and 0.206,
respectively.

The Stanford Web data can be decomposed into 29,914 SCCs. Among
these, 26,396 components have size 1, meaning that they are single
nodes; the other 3,518 components consist of two or more nodes.
The largest SCC contains 150,532 nodes-- 53.4\% of the total
number of nodes in the network. We note that the Stanford data
does not comprise a single connected network. Instead it contains
365 WCCs, the biggest of which consists of 255,265 nodes, or
90.6\% of the total nodes in the network. The size distributions
of the SCCs and the WCCs are displayed in the
Fig.~\ref{fig:StanfordDegree}(c) and (d).

Figure~\ref{fig:SCCdiagram} portrays the SCC structure of the 98
biggest SCCs in the Stanford Web data. For better visualization,
only the 98 largest SCCs and their connecting links have been
depicted. The smallest (98th) SCC in this diagram contains 162
nodes, and altogether, these 98 SCCs contain 198,123 nodes, which
corresponds to 70.3\% of the total nodes in the network. The size
of each circle in Fig.~\ref{fig:SCCdiagram} maps to the size of
the corresponding SCC, and the color indicates whether a SCC lies
inside or outside the giant WCC (blue is inside, and red is
outside). As can be seen in Fig.~\ref{fig:SCCdiagram}, the network
clearly exhibits a simple {\em bow-tie}
structure~\cite{Broder2000}.

\end{document}